\documentclass[sn-basic]{sn-jnl-revised}

\usepackage[utf8]{inputenc}
\usepackage{subfigure}
\usepackage{threeparttable}
\usepackage{booktabs}
\usepackage{amsmath}
\usepackage{verbatim}
\usepackage{longtable}
\UseRawInputEncoding
\jyear{2022}%

\theoremstyle{thmstyleone}%
\theoremstyle{thmstyletwo}%

\theoremstyle{thmstylethree}%

\begin{document}
	
	\title[Return migration in academia]{Return migration of German-affiliated researchers: Analyzing departure and return by gender, cohort, and discipline using Scopus bibliometric data 1996-2020}

	\author[1,2]{\fnm{Xinyi} \sur{Zhao}}
	
	\author*[3,1]{\fnm{Samin} \sur{Aref}}\email{aref@mie.utoronto.ca}
	
	\author[1]{\fnm{Emilio} \sur{Zagheni}}
	\author[4]{\fnm{Guy} \sur{Stecklov}}
	
	\affil[1]{\orgdiv{\footnotesize Lab of Digital and Computational Demography}, \orgname{Max Planck Institute for Demographic Research}, \orgaddress{\street{Konrad-Zuse-Str. 1}, \city{Rostock}, \postcode{18057}, \state{Mecklenburg-Vorpommern}, \country{Germany}}}

	\affil[2]{\orgdiv{\footnotesize Leverhulme Centre for Demographic Science, Department of Sociology}, \orgname{University of Oxford}, \orgaddress{\street{42-43 Park End Street}, \city{Oxford}, \postcode{OX1 1JD}, \country{UK}}}
	
	\affil[3]{\orgdiv{\footnotesize Department of Mechanical and Industrial Engineering}, \orgname{University of Toronto}, \orgaddress{\street{5 King's College Rd}, \city{Toronto}, \postcode{M5S 3G8}, \state{ON}, \country{Canada}}}
	
	\affil[4]{\orgdiv{\footnotesize Department of Sociology}, \orgname{University of British Columbia}, \orgaddress{\street{2329 West Mall}, \city{Vancouver}, \postcode{V6T 1Z4}, \state{BC}, \country{Canada}}}

	\abstract{
		The international migration of researchers is an important dimension of scientific mobility, and has been the subject of considerable policy debate. However, tracking the migration life courses of researchers is challenging due to data limitations. In this study, we use Scopus bibliometric data on eight million publications from 1.1 million researchers who have published at least once with an affiliation address from Germany in 1996-2020. We construct the partial life histories of published researchers in this period and explore both their out-migration and the subsequent return of a subset of this group: the returnees. Our analyses shed light on the career stages and gender disparities between researchers who remain in Germany, those who emigrate, and those who eventually return. We find that the return migration streams are even more gender imbalanced, which points to the need for additional efforts to encourage female researchers to come back to Germany. We document a slightly declining trend in return migration among more recent cohorts of researchers who left Germany, which, for most disciplines, was associated with a decrease in the German collaborative ties of these researchers. Moreover, we find that the gender disparities for the most gender imbalanced disciplines are unlikely to be mitigated by return migration given the gender compositions of the cohorts of researchers who have left Germany and of those who have returned. This analysis uncovers new dimensions of migration among scholars by investigating the return migration of published researchers, which is critical for the development of science policy.}
	
	
	\maketitle
	
	
	\section{Introduction}\label{sec1}
	
	To ensure the dynamic flow of scientific ideas and expertise, and to promote and facilitate knowledge production, national science systems rely on the international exchange of scholars \citep{moed_studying_2013,conchi_scientific_2014,robinson-garcia_many_2019}. While the globalization of research has numerous benefits that have been widely acknowledged in the literature \citep{appelt_chapter_2015,bauder_international_2015,franzoni_chapter_2015,netz2017explaining}, the undeniable downside of academic mobility is the potential loss of talent for countries that train and export more researchers than they receive from other countries. Global competition for talent has led to the introduction of a range of policies and economic incentives aimed at encouraging balanced flows of researchers. However, little attention has been paid to returnees who stay in other countries temporarily, and then return to their country of academic origin. These returnees usually bring with them additional skills, newly established connections and collaborative ties, and complementary expertise acquired abroad \citep{oecd_global_2008, appelt_chapter_2015}. Equally importantly, there is evidence that these returnees tend to receive far more citations than their stationary counterparts \citep{guthrie_international_2017}, or their internationally mobile counterparts who do not return \citep{zhao2021international}. Thus, for countries that are facing the challenge of the out-migration of researchers exceeding the in-migration of researchers \citep{zhao2021international}, facilitating the return migration of scholars and taking steps to rebalance these trends to their benefit are extremely critical.
	
	Despite being recognized as a science powerhouse, Germany has been sending more highly qualified individuals (including researchers) abroad than it has been receiving according to some reports \citep{oecd_global_2008, schiller_measuring_nodate,zhao2021international}. In recent years, Germany has implemented a range of policies and programs designed to attract students and researchers from other countries \citep{al_problem_2016,eule_inside_2016,duvell_germany_nodate}. There are several return migration programs aimed at maintaining and strengthening ties with previously German-affiliated researchers in order to facilitate their re-integration into the German science system \citep{conchi_scientific_2014}. For example, the German Academic International Network (GAIN)\footnote{\url{www.gain-network.org}} supports current and prospective returnees, and facilitates cooperation between researchers in Germany and North America. The German Scholars Organization (GSO)\footnote{\url{www.gsonet.org}} is another initiative aimed at reversing Germany's ``brain drain" and turning it into a ``brain gain" by offering several services to academic professionals in Germany. Given the practical relevance of this issue to policy development and strategic decisions at a national level, a better understanding of the trajectories and migration trends of researchers formerly affiliated with German institutions and academic returnees to Germany is urgently needed.
	
	Return migration also has potentially large implications for the persistence of gender inequalities in academia \citep{zippel_women_2017}. The issue of gender disparities in academia has been extensively documented across many disciplines and in most countries \citep{lariviere2013bibliometrics,huang_historical_2020,zhao2021international}. Although it has been suggested that breaking the glass ceiling that hinders women's advancement is especially challenging for internationally mobile academics \citep{zippel_women_2017,zhao2021international}, the heterogeneity in the levels of gender disparity among former and current German-affiliated researchers, and particularly among returnees, is not clear. Exploring this topic is the first step towards achieving more balanced gender representation in academia, and ensuring the sustainability of academic careers for women researchers \citep{weert_support_2013,zhao2021international}. Due to the implementation of a wide range of measures and policies aimed at promoting gender equality in academia, female representation in various disciplines have been increasing \citep{macaluso_is_2016,zippel_women_2017,huang_historical_2020}. Given these developments, studying the temporal trends at the intersection of gender and international mobility can have important administrative and policy implications. Moreover, the question of how the representation of female returnees in more recent cohorts has changed in response to these policies and developments deserves more attention. An extensive analysis of return migration among German-affiliated researchers disaggregated by discipline, gender, and cohort is critical for making progress towards transforming the German science system into an inclusive and diverse system with balanced migration flows of scholars.
	
	Previous studies have suggested that academic returnees tend to maintain collaborative ties with their previous host countries \citep{ackers_attracting_2005, conchi_scientific_2014,franzoni2014mover,guthrie_international_2017}. As these ties are critical components of knowledge transfers, it is clear that scholarly migration is no longer a zero-sum game. Returnee researchers may face challenges when attempting to incorporate the knowledge they have acquired abroad into another context, and in re-establishing their career in their academic home country \citep{melin_what_2006,weert_support_2013,fernandez-zubieta_chapter_2015}. This may be partly because researchers are disconnected from their home country's academic networks while abroad, which may limit their access to the information and support they would need to find a job in their academic home country. This erosion of connections may, in turn, reduce the willingness of researchers to return \citep{ackers_attracting_2005, baruffaldi_return_2012}. Previous work on this topic has shown that there is a gap in macro-level quantitative research on collaboration and migration among scholars. In particular, the interactions between scholars' collaborative ties with their academic home countries and return migration have not been previously investigated. Therefore, a comprehensive analysis that examines these relationships can provide useful insights into return migration among academics. This can facilitate the development of policies that create additional paths for returnees to re-integrate professionally into their academic home countries.
	
	Motivated by the observations above, this paper relies on large-scale digitized bibliometric data from Scopus \citep{burnham_scopus_2006, mongeon_journal_2016} to investigate the trajectories and migration trends of internationally mobile researchers in Germany as well as their German academic links during the period of being outside of Germany. We analyze German-affiliated published researchers during the 1996-2020 period, while taking each researcher's years of experience, gender, discipline, and cohort into account. Specifically, this paper aims to address the following research questions (which are both methodological and empirical), with a focus on the return migration of scholars to Germany:
	\begin{enumerate}
		\item\label{th:first} What is the composition of returning researchers based on their gender, years of experience, and previous host countries? (Subsection \ref{ss:Pyramids} and \ref{ss:Geo})
		\item \label{th:second} How does the gender ratio vary by discipline and cohort among researchers who leave Germany, and among researchers who return? (Subsection \ref{ss:gender} and \ref{ss:discipline})
		\item \label{th:third} How does the association between return migration to Germany and collaboration with German institutions vary across disciplines and cohorts? (Subsection \ref{ss:collaboration})
	\end{enumerate}
	
	\section{Materials and methods}\label{sec2}
	
	\subsection{Authorship records of German-affiliated researchers}\label{subsec1}
	Scopus is an abstract and citation database of scientific literature \citep{burnham_scopus_2006} that covers over 77 million publications, according to its 2020 coverage guide \citep{scopus_coverage}. From the Scopus database, we have obtained the authorship records (linkages between an author's affiliation and a publication) of more than 1.1 million researchers for our analysis, which involves over eight million publications. All of these researchers had used a German affiliation address in at least one publication at some point during the 1996-2020 period (ending in April 2020).
	
	\subsection{Pre-processing of raw bibliometric data}\label{subsec2}
	To ensure the reliability of our results, we pre-process the raw bibliometric data from Scopus using three sequential steps. These three steps are: handling the missing countries in the dataset (discussed in subsection \ref{ss:missing}), disambiguating the author profiles (discussed in subsection \ref{ss:disambiguation}), and inferring gender from the authors' first names (discussed in subsection \ref{sss:gender}).
	
	\subsubsection{Handling missing countries in the dataset}\label{ss:missing}
	
	First, there are 74,430 (5\%) authorship records for which the country variable is missing. To handle the missing values systematically, we have developed a neural network algorithm inspired by \cite{gonzalez2020scholarly} that predicts the missing countries with a high degree of accuracy. This supervised learning algorithm takes an affiliation address as the input, and predicts the country as the output. The data used to make the prediction are city, institution, and address. These strings are combined using a bag-of-words method with frequencies (of a given word in a sample) that are normalized (relative to the frequency in the whole dataset) using a \textit{term frequency inverse document frequency} (tfidf) approach \citep{tfidf}. A simple and standard architecture is used to develop the neural network (deep feed forward neural network) with non-linear activation functions. We use a random set of one million authorship records drawn from our dataset that contain country information and split it into training data (80\%) and testing data (20\%). Other technical details of the development of the neural network have been explained in \citep{gonzalez2020scholarly}. The predictions made based on the test dataset show that the neural network can correctly predict the country for 98.4\% of records, which is a level of accuracy we consider acceptable for handling missing country data.
	
	\subsubsection{Author name disambiguation}\label{ss:disambiguation}
	The second step of our data pre-processing helps us overcome the problems associated with using Scopus author IDs to identify unique authors. It has been shown that Scopus author IDs have high levels of precision and completeness \citep{kawashima_accuracy_2015,loktev2020}. Precision measures the percentage of author IDs that are associated with the publications of a single individual only. Completeness measures the percentage of author IDs that are associated with all of the Scopus publications of an individual. The results of an evaluation of the accuracy of Scopus author IDs conducted in August 2020 showed that the precision and the completeness of Scopus author profiles are 98.3\% and 90.6\%, respectively \citep{loktev2020}. However, while it appears that the quality of individual-level Scopus data is sufficiently high to enable us to study the migration of researchers \citep{kawashima_accuracy_2015, aman_does_2018}, there are several notable limitations to keep in mind when using Scopus data for migration research. The precision limits in Scopus author IDs imply that 1.7\% of Scopus author IDs may be associated with the publications of more than one person, which could affect the accuracy of the migration events detected by looking at changes in affiliation countries per author ID. Accordingly, as the second step in our data pre-processing, a subset of authorship records that are more likely to have suffered from the precision flaws of Scopus author IDs are analyzed using our conservative author name disambiguation algorithm.
	
	Our author disambiguation algorithm is inspired by the state-of-the-art methods in author name disambiguation \citep{dangelo_collecting_2020}. It assumes that every two authorship records are from distinct individuals (despite sharing a Scopus author ID), unless sufficient evidence is found to the contrary using a rule-based scoring approach and a clustering method. We first calculate the similarity score of each pair of authorship records belonging to the same author ID. The similarity is measured based on author names, coauthor names, subjects, funding information, and grant numbers. The author disambiguation algorithm makes all pairwise comparisons between authorship records with the same author ID, and creates a distance matrix based on similarities and dissimilarities in the aforementioned features for each pair of records. A clustering algorithm is then used to process the distance matrices, and to cluster similar authorship records. We then issue revised author IDs based on the resulting clusters. We use the agglomerative clustering algorithm from the scikit-learn Python library \citep{pedregosa_scikit-learn_2011} to cluster authorship records. This algorithm belongs to the family of hierarchical clustering methods. Supporting our conservative approach, it first places each record in its own cluster, and then merges pairs of clusters successively if doing so minimally increases a given linkage distance \citep{pedregosa_scikit-learn_2011}. As well as being compatible with our conservative approach, agglomerative clustering has the advantage of offering us the flexibility to process any pairwise distance matrix.
	
	We examine the author profiles that are outliers in terms of the number of affiliation countries or the number of publications. In particular, there are 30,715 (2\%) author profiles that are associated with more than six countries of affiliation, or more than 292 publications\footnote{These two thresholds are chosen empirically so that a subset of outliers of a size compatible with the results on Scopus author ID precision flaws \citep{loktev2020} can be extracted.} (an average of more than one publication per month across a period of 24 years and four months). These author profiles are more likely than others to be affected by the precision flaws of Scopus author IDs. For example, each Scopus author profile could contain records from more than one individual researcher. Based on these criteria, 25,000 author IDs are classified as suspicious. These author IDs are associated with 2,242,797 publications. After disambiguation, revised author IDs are issued for these records according to their clusters, and are then merged with the rest of the data in preparation for the third pre-processing step.
	
	\subsubsection{Inferring gender from first names}\label{sss:gender}
	
	The last step of our data pre-processing is inferring gender from first names, which involves looking up first names in a large database of names and genders called \textit{Genderize} (\url{genderize.io}). After performing basic text operations (like removing middle initials from the first name field), we obtained the gender for 1,117,813 author profiles in our dataset. For the remaining profiles, we manually searched for public author information to determine the gender by checking the individuals' personal homepages, curricula vitae, online profiles, and biographies in publications, as well as other online sources. Using this manual approach, we were able to determine the genders for 3,139 additional author profiles. Finally, the most likely gender for 77\% of the author profiles in our dataset was determined through either algorithmic or manual gender detection. For our analyses that involve gender (e.g., measuring gender ratios), we set aside the 23\% of author profiles whose gender could not be determined either algorithmically or manually.
	
	\subsection{Migration events, mobility types, and career stages}
	\label{subsec3}
	
	The international mobility of researchers is determined by identifying the changes in the affiliation addresses of authors across different publications over time. To more reliably detect migration events, the most frequent (mode) country(ies) of affiliation is extracted for each researcher in each year. A migration event is considered to have happened only if the mode country of affiliation changes for the researcher across different years \citep{subbotin2020brain}. Accordingly, the \textit{country of academic origin} (\textit{country of academic destination}) is defined as the mode country during the first (last) year of publishing. Based on the individual's migration events or the lack thereof, each researcher can be assigned to one of the following four categories:
	\begin{enumerate}
		\item  Non-mover (with Germany being the researcher's mode country in all years); 
		\item  Immigrants and transients (origin: not Germany; but with Germany being the researcher's mode country at some point in time); 
		\item  Outward (origin: Germany; current country: not Germany); and
		\item  Returnee (origin and current country: Germany; but with another country being the researcher's mode country at some point in time).
	\end{enumerate}
	
	Except for non-movers, researchers may move between the categories over time, as an individual's status depends on the time period being examined. For example, an individual identified as an outward researcher will become a returnee at the next point in time if a move to Germany is detected.
	
	We define the \textit{academic age} (\textit{age}) of a researcher as the number of years since his/her first publication. Furthermore, we classify researchers as \textit{early-career} (\textit{senior}) if their academic age is seven years or less (14 years or more). Researchers with an academic age between seven and 14 years are categorized as \textit{mid-career} \citep{aref_demography_2019}. As our dataset covers only the 1996-2020 period, our analysis of some temporal dimensions of the data or cohorts of researchers may suffer from left truncation and/or right censoring. We explain in Section \ref{sec:result} some of the resulting limitations of our dataset.

	\subsection{Inferring disciplines using a topic model}\label{subsec4}
	The Science Journal Classification (ASJC) codes in our bibliometric dataset indicate the fields and disciplines (subfields) of publication venues, which could be used as proxies for determining the disciplines of researchers \citep{zhao2021international}. However, because the links between the disciplines associated with journals and the disciplines of authors are indirect, we use a data-driven method to infer the disciplines of individual researchers. Topic modelling, which is a common unsupervised learning approach for natural language processing, can be used to determine the disciplines of researchers by inferring the latent topical structure of textual bibliometric data \citep{blei_probabilistic_2012, gerlach_network_2018}.
	
	As a flexible topic model, Latent Dirichlet Allocation (LDA) is in essence a generative probabilistic model with three layers: document, topic, and word  \citep{pritchard_inference_2000, blei_latent_2003}. It assumes that each topic is a mixture of an underlying set of words, and that each document is a mixture of a set of topic probabilities \citep{blei_latent_2003, gerlach_network_2018}. LDA has been shown to perform reliably in automatically identifying semantic topic information from large-scale textual data \citep{dahal_topic_2019}.
	
	From the publications authored by each researcher in our dataset, we extract publication titles, journal titles, and keywords to generate the individual's text corpus (document in LDA terminology). We then tokenize the text by removing all punctuation, and making all of the words lower-case to improve the cohesion of the documents. The remaining words in each text corpus are then lemmatized and stemmed to their root form. This includes being transferred to the first person and the present tense if needed. For some common phases (e.g., machine learning) that are related to discipline topics, we use the multi-word expressions, two-gram collocations, and three-gram collocations from all of the text documents according to their frequency of occurrence. The tokenized and lemmatized texts are then abstracted to a \textit{bag of words}, which records the indices of words and the number of times each word appears in an author's LDA document. In LDA, each document can be considered as a mixture of latent topics, each of which is characterized by a distribution of words \citep{blei_latent_2003}. The topic coherence score is a measurement of the semantic similarity between the high scoring words in each topic, and represents the interpretability of the topics.

	After the implementation of the steps above, the average topic coherence score of all topics is maximized at $0.67$ (through trial and error) when we allow 30 topics for the whole text corpus
	. Each topic is composed of a set of vocabularies and the corresponding weights that indicate their contributions to the topic. The topic with the highest probability among all of the initial topics is assigned to each author's LDA document as the intermediate result. We manually interpret the topics based on their most frequent terms and assign titles to them accordingly. When different topics include the similar or highly relevant keywords, we combine them into a single discipline. For example, a topic involving ``space" and ``earth" and a topic involving ``galaxy" and ``star" are combined to the same discipline: ``Earth and Planetary Sciences." Using this approach, we produce 17 distinguishable disciplines\footnote{``Agricultural, Biological and Environmental Sciences," ``Biochemistry, Genetics and Molecular Biology," ``Chemistry and Chemical Engineering," ``Computer Science," ``Earth and Planetary Sciences," ``Economics and Social Science," ``Engineering," ``Energy," ``Health Professions," ``Immunology and Microbiology," ``Materials Science," ``Mathematics," ``Medicine," ``Neuroscience," ``Pharmacology, Toxicology and Pharmaceutics," ``Physics and Astronomy," and ``Psychology."} to represent the main discipline of each researcher according to their publications. Detailed results on the 30 topics and their mapping to 17 disciplines are provided in the appendix. We consider an author's LDA document to be ``Multidisciplinary" if it does not have any contribution percentage for any topic that exceeds $0.3$. 
	
	\subsection{Cohorts leaving Germany and returning to Germany}\label{subsec5}
	A cohort is a group of people who have experienced a common event in a selected period, such as birth \citep{reilly_early_2005,rothman_epidemiology_2012}. In our analysis, we use the time of first publication as the common event for defining cohorts of researchers. To reduce the impact of left-truncated data on cohorts, which is more likely for the first few years of our dataset, we use the following three cohorts: 1998-2001, 2002-2005, and 2006-2009.
	
	\textit{Person-time rate} is an index commonly used in epidemiology and demography to express an incidence rate: i.e., the number of incidents (migration events) per person-time in a population during a period \citep{rothman_epidemiology_2012}. The denominator of a person-time rate is the total amount of time that the study members are at risk of a certain incident during a period. One key advantage of using person-time rate for migration is that it enables us to consider that different individuals are exposed to migration events for varying amounts of time. The incidents we are interested in are: (1) leaving Germany for the group of all researchers in Germany, and (2) returning to Germany for the group of all outward researchers. Given a specific period of time $t$, the \textit{departure rate} per 1,000 person-years for a cohort $c$ is defined in Eq.\ \eqref{eq1}.
	
	\begin{equation}
		\label{eq1}
		R_{\text{departure}(c,t)}=N_{\text{departure}(c,t)}/ PT_{\text{in Germany}(c,t)} \times 1000
	\end{equation}
	
	In Eq.\ \eqref{eq1}, $N_{\text{departure}(c,t)}$ represents the number of researchers from cohort $c$ leaving Germany during time period $t$, and $PT_{\text{in Germany}(c,t)}$ represents the sum of the number of years each researcher from cohort $c$ stays in Germany (and is exposed to leaving Germany) during period $t$. The denominator of the departure rate takes all of the researchers who are in Germany into consideration as the population exposed to leaving Germany.
	Similarly, given a specific period of time $t$, the \textit{return rate} per 1,000 person-years for a cohort $c$ is defined in Eq.\ \eqref{eq2}.
	
	\begin{equation}
		\label{eq2}
		R_{\text{return}(c,t)}=N_{\text{return}(c,t)}/ PT_{\text{outside Germany}(c,t)}\times 1000
	\end{equation}
	
	In  Eq.\ \eqref{eq2}, $N_{\text{return}(c,t)}$ is the number of returnees from cohort $c$ during period $t$, and $PT_{\text{outside Germany}(c,t)}$ is the sum of the number of years each outward researcher stays outside of Germany (and is exposed to returning to Germany) during period $t$. The denominator of the return rate only involves researchers who have left Germany as the population exposed to returning to Germany.
	We compute the departure rates and the return rates separately for male and female researchers in the three cohorts (1998-2001, 2002-2005, and 2006-2009). Specifically, we consider the departure rates of researchers of different cohorts who leave Germany at the academic ages of one to five, and the corresponding return rates during the first five years after their departure from Germany. Taking the 1998-2001 cohort as an example, the departure rate at academic age one refers to the outward researchers who were ``academically born" in 1998 (1999, 2000, 2001) and leave Germany in 1999 (correspondingly, 2000, 2001, 2002). For the same cohort, the return rate refers to the researchers who returned during the 2000-2004 (2001-2005, 2002-2006, 2003-2007) period among the outward researchers who left Germany in 1999 (correspondingly, 2000, 2001, 2002).
	
	\subsection{Collaborations with Germany while away}\label{subsec6}
	
	We define and use the variable \textit{collaborative ratio} to distinguish the strength of the academic linkages with Germany for outward researchers during the period when they were away from Germany. For the outward researcher, $i$, during the period when s/he was away from Germany (denoted by $t$), we calculate his/her collaborative ratio using a simple fraction: $CR_{(i)}=D_{(i,t)}/ N_{(i,t)}$. The numerator, $D_{(i,t)}$, counts the publications of outward researcher $i$ in period $t$ that list a German affiliation for $i$ or for his/her co-authors. The denominator, $N_{(i,t)}$, is the number of all publications of outward researcher $i$ during period $t$. If a publication authored by $i$ during period $t$ has at least one author with a German affiliation, it contributes to the collaborative ratio $CR_{(i)}$. Furthermore, the average collaborative ratio for all researchers in each discipline (cohort) is calculated to measure the average strength of the academic collaboration with Germany maintained by the outward researchers in that discipline (cohort).
	
	\section{Results}\label{sec:result}
	
	Using cleaned and processed bibliometric data associated with over eight million Scopus-indexed publications over 1996-2020 from more than one million German-affiliated researchers, we analyze data on 375,288 female researchers associated with 2,665,139 publications and 745,664 male researchers associated with 6,516,016 publications. Among these researchers, there are 50,803 female mobile researchers (associated with 1,007,606 publications) and 119,298 male mobile researchers (associated with 2,760,282 publications) who have ever migrated between Germany and 194 other countries in our dataset. There are $103,573 \pm 48610$ researchers in each discipline, with medicine having the largest number of researchers (199,658) and health professions having the smallest number of researchers (26,398).
	
	Based on pre-processed data, we provide five analyses to describe different aspects of the emigration and the return migration of these researchers. In Subsections \ref{ss:Pyramids} and \ref{ss:Geo}, we track their career life courses from a temporal perspective, and their geographic trajectories from a spatial perspective. We then compare the departure rates and the return rates of female and male scholars to explore the gender differences across cohorts and disciplines (Subsections \ref{ss:gender} and \ref{ss:discipline}). Finally, we look at the association between the return rates of outward researchers and the strength of their collaborative ties to Germany in Subsection \ref{ss:collaboration}. Our inferred migration events dataset is publicly available in a FigShare data repository \citep{zhao2022data}.
	
	\subsection{Age and gender composition of researchers} \label{ss:Pyramids}
	
\begin{figure}[!htbp]
		\centering
		\vspace{-0.35cm} 
		\subfigtopskip=2pt 
		\subfigbottomskip=2pt 
		\subfigcapskip=-5pt 
		\subfigure[Non-movers]{
			\label{fig:1.sub.1}
			\includegraphics[width=0.33\linewidth]{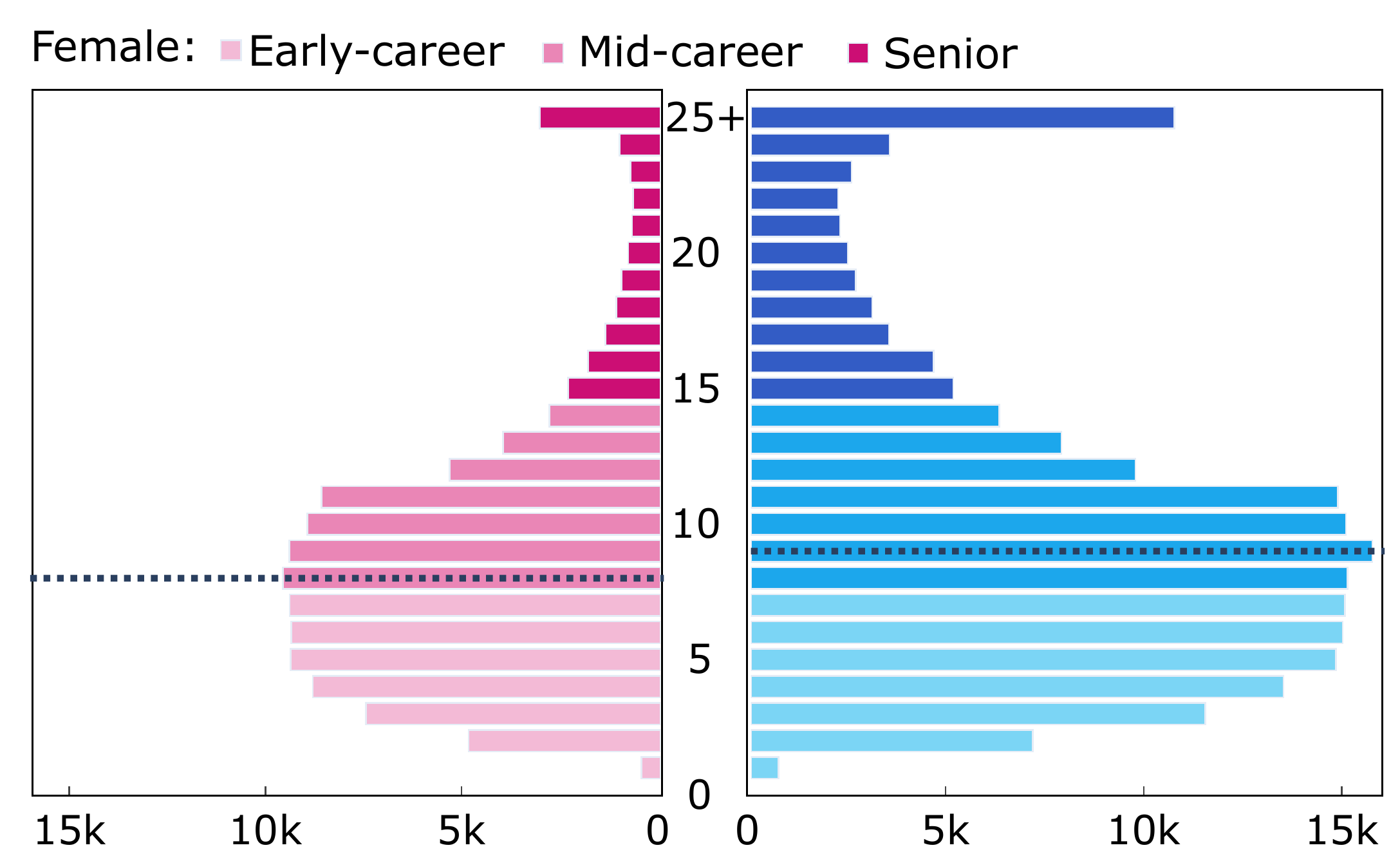}}
		\subfigure[Outward researchers]{
			\label{fig:1.sub.2}
			\includegraphics[width=0.335\linewidth]{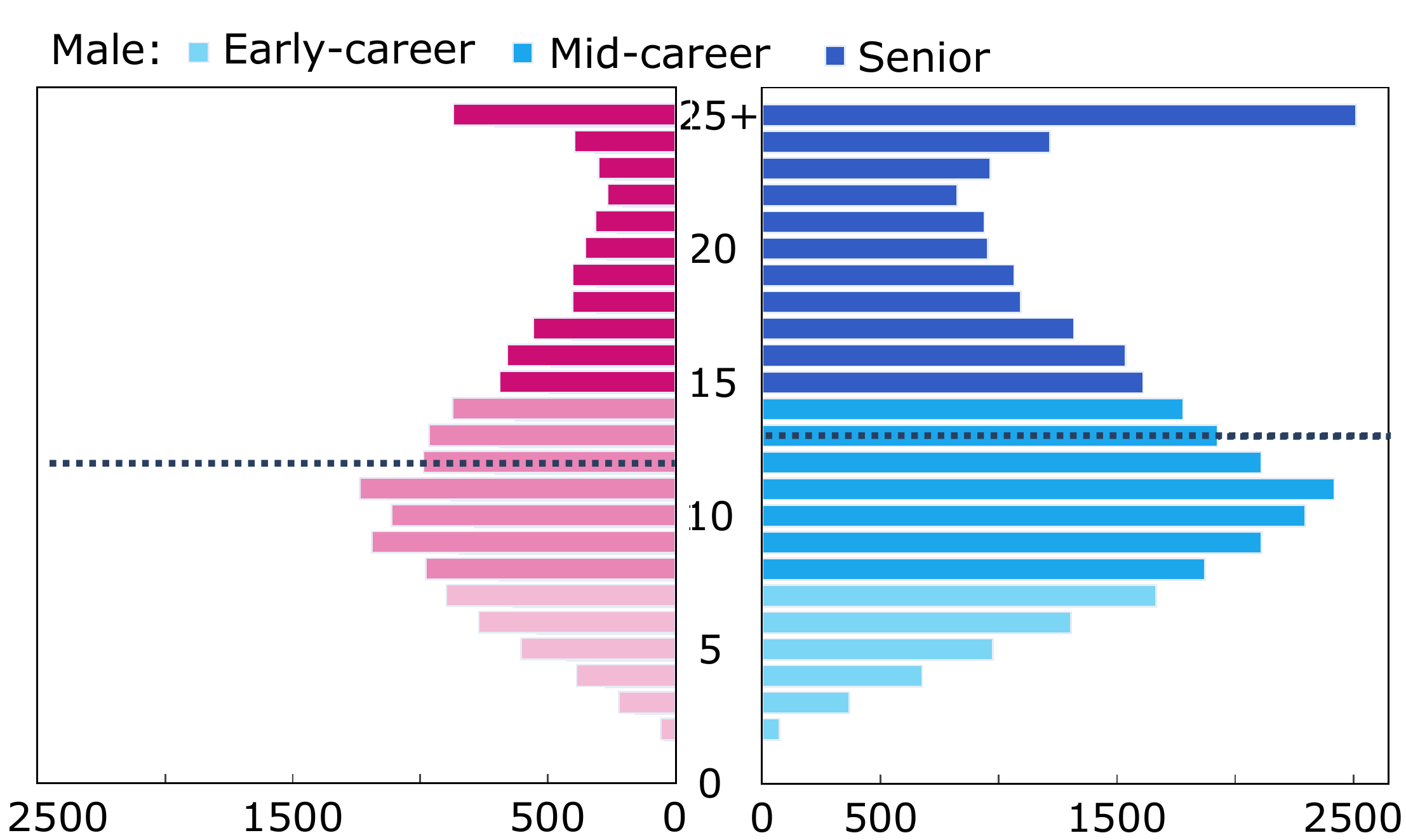}}
		\subfigure[Returnee researchers]{
			\label{fig:1.sub.3}
			\includegraphics[width=0.258\linewidth]{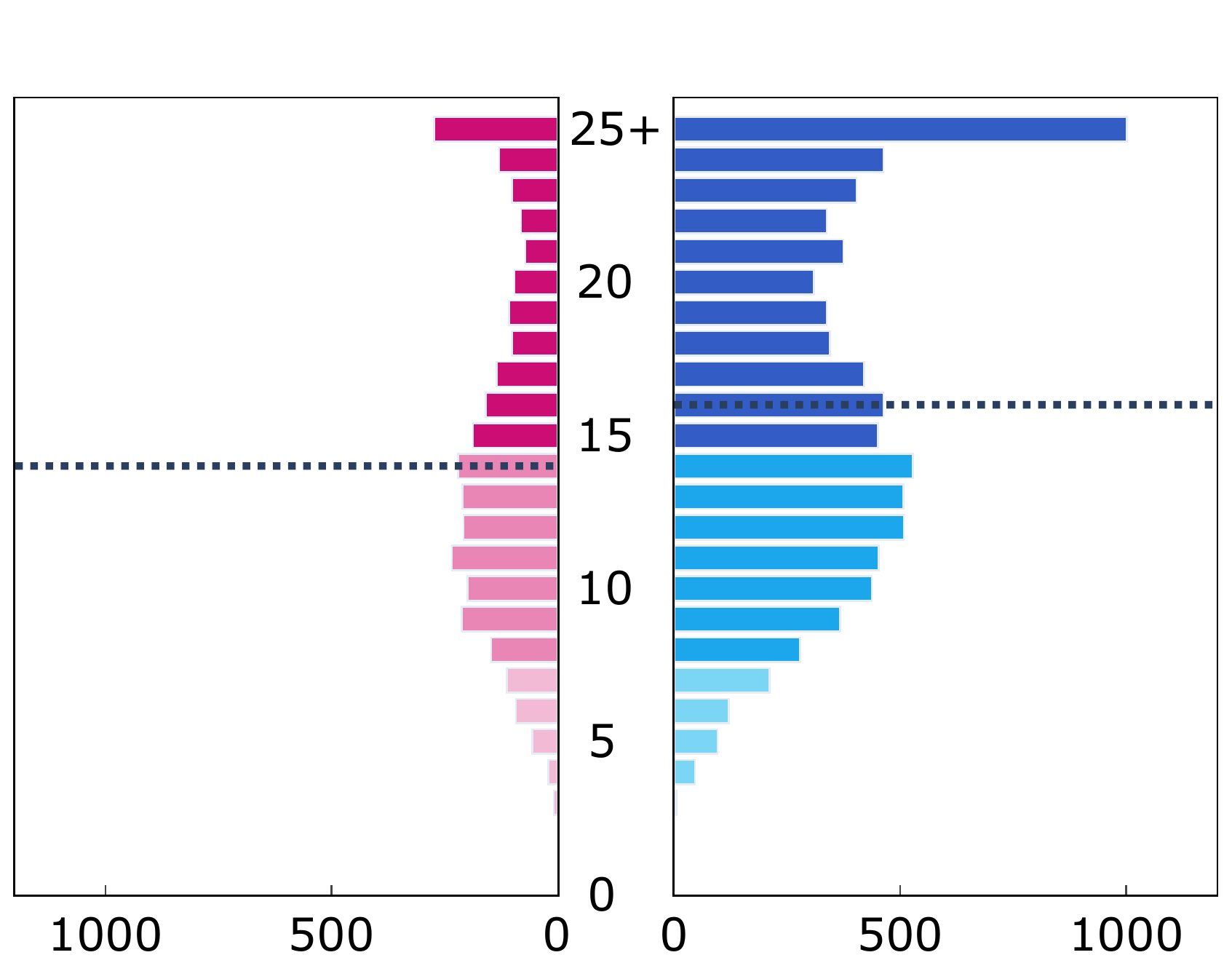}}
		\caption{Composition of academic age and gender for non-movers, outward researchers, and returnees} 
		\label{fig:1}
	\end{figure}

	We compare the age and gender compositions of three groups of researchers: non-movers, outward researchers, and returnees. Figure \ref{fig:1} compares the age and gender distribution of these three groups using population pyramids, which include individuals who survive as an active researcher up to a certain date (researchers whose latest publication was in 2010 or later). Ignoring the truncated top age, we can see that both female and male non-movers have a notable and pronounced bulge at the transition from early-career ages to middle-career ages, which is presented as an expansive pattern. However, the non-mover age pyramid shows considerably small proportions at other ages, with a pattern characterized by a sharp decline to age 21, followed by a stable increase until age 25+. The median ages for female and male non-movers are nine and 10 years, respectively. In the categories of outward researchers and returnees, the overall length of academic trajectories has been considerably prolonged for both female and male researchers. Specifically, the median ages of female outward researchers and returnees are 12 and 14, respectively; and the median ages of male outward researchers and returnees are 13 and 16, respectively.
	
	Overall, these findings suggest that both the male and the female returnees stayed in academia longer than both the outward researchers and the non-movers. As more evidence on return migrants emerges, the strengths of returnees are increasingly being seen as valuable. It has, for example, been shown that from a historical perspective, returnees tend to make important contributions to local economies and to be relatively successful, both in comparison to people who never migrated and to people who emigrated but did not return \citep{abramitzky2019new}. The findings on the positive impact of return migration are encouraging, and suggest that Germany, as well as other sending countries, should embrace international mobility and the return migration of scholars. While we find that both the male and the female scholars benefited from returning, we also observe that returning to Germany had a more positive impact on the careers of male than of female researchers. For example, the results show that 64.57\% of male returnees, but only 51.13\% of female returnees, had become senior professionals (see detailed age composition in \ref{fig:1.sub.3}). The smaller benefits found for women are not surprising, and point to the ongoing challenges women in academia face.
	
	\subsection{Out-migration and return migration by geography}\label{ss:Geo}
	
\begin{figure}[!htbp]
		\centering
		\includegraphics[width=0.95\textwidth]{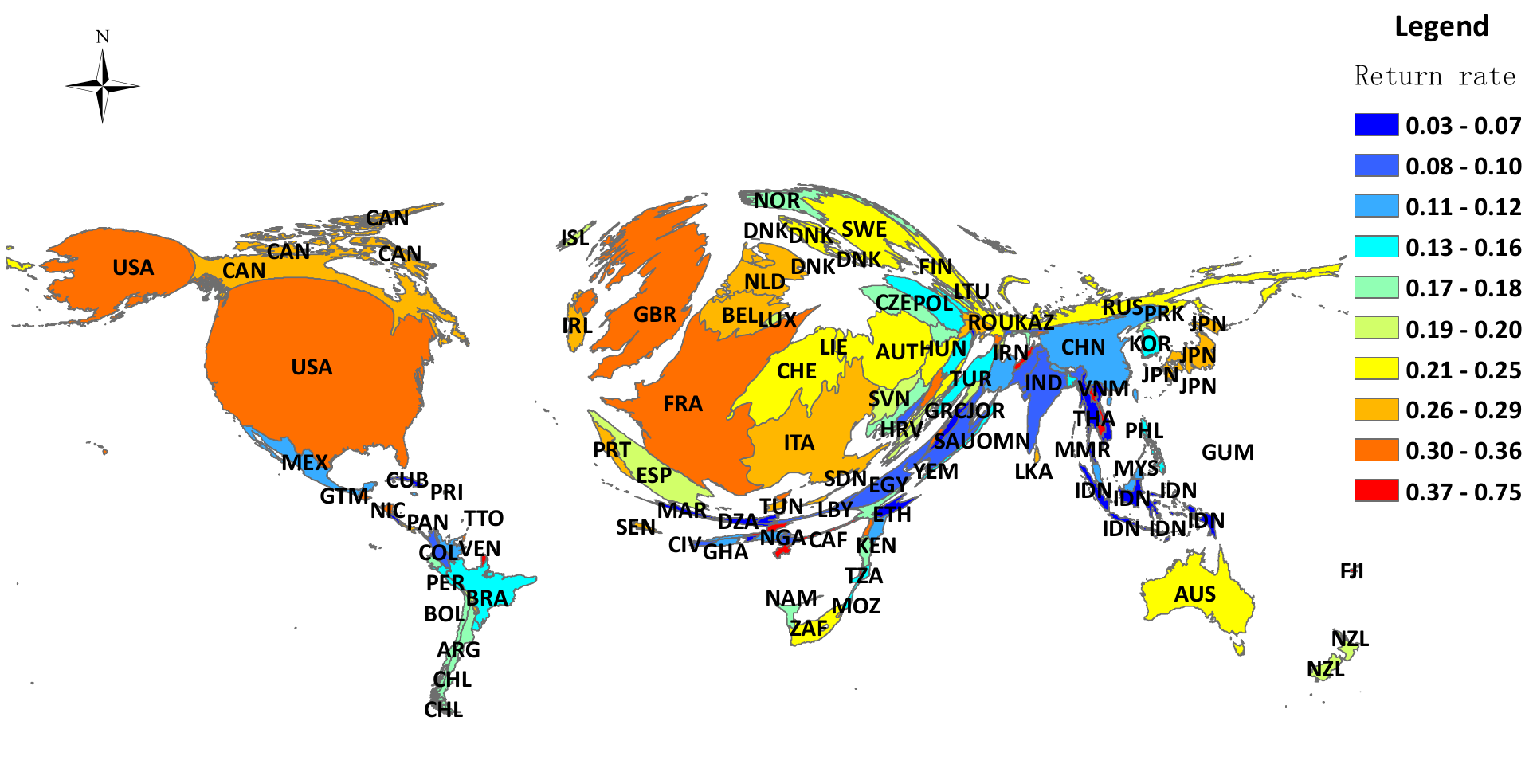}
		\caption{Outward flows (from Germany) and respective return rates across countries. The sizes of the countries are proportional to the flows of outward researchers from Germany. The colors indicate the differences in the return rates of the German-affiliated researchers returning to Germany from each country.}
		\label{fig:2}
	\end{figure}

	Figure \ref{fig:2} illustrates from a geographic perspective the interplay between outflows of researchers from Germany and the corresponding rates of return to Germany, through a \textit{density equalizing cartogram} \citep{dorling_worldmapper_2006,houle_use_2009}. Here, the shape of the map polygons is transformed proportionally to the outflows to different countries. The colors represent the differences in the countries' return rates, as further explained in the legend. The most common host country for researchers from Germany was the United States (US), which received around 24\% of the outward researchers from Germany. Next came Switzerland and the United Kingdom (UK), which together attracted 22\% of the outward researchers from Germany. In total, these three countries received nearly half of the outward researchers from Germany, and had thus become the most appealing options for German researchers interested in pursuing an international academic career. These estimates are also consistent with previous findings that the US, the UK, and Switzerland are the most common origin and destination countries for scholarly migration to and from Germany \citep{OECD_return_2015,zhao2021international}. The observed pattern for the European countries that received researchers from Germany indicates that the countries that neighbor Germany and German-speaking countries were among the most popular host countries for scholars who began their publishing activity in Germany.

	As the colors on the map show, the rates of return from the most common receiving countries that are larger in size were all below 36\%; meaning that about one-third of German-affiliated researchers moved back to Germany, while nearly two-thirds continued their research abroad. While the US hosted the largest share of researchers from Germany, the rate of return to Germany from the US was also relatively high, at 34\%. Similarly, while the UK and France were among the top host countries for researchers from Germany, the rates of return to Germany from these countries were also high, at 30\% and 29\%, respectively. By contrast, the rates of return to Germany were below one-quarter for German-affiliated researchers in Switzerland, Sweden, Austria, and Australia; and the return rate was especially low for German researchers in Switzerland, at only 20\%. It thus appears that researchers who moved from Germany to these four countries were comparatively less likely to return. The lower propensity to return may be partly explained by the higher spending on Research \& Development (R\&D) in these countries. In 2017, Switzerland, Sweden, Austria, and Australia spent about 3.18\%, 3.36\%, 3.00\%, and 3.08\% of their GDP, respectively, on R\&D – far above the OECD average of 2.67\%, ahead of the US (2.85\%), the UK (1.68\%), and on levels competing with Germany (3.07\%) \citep{RDGDP}. In addition, the lower return rate of German researchers in Switzerland is broadly consistent with our expectations, given that approximately 1.2\% of all scientific papers worldwide are produced by Swiss-affiliated researchers, which is remarkable given the country's small population \citep{ResearchinSwitzerland}.

	\subsection{Rates of departure and return across cohorts}
	\label{ss:gender}
	
\begin{figure}
		\centering
		\includegraphics[width=0.9\textwidth]{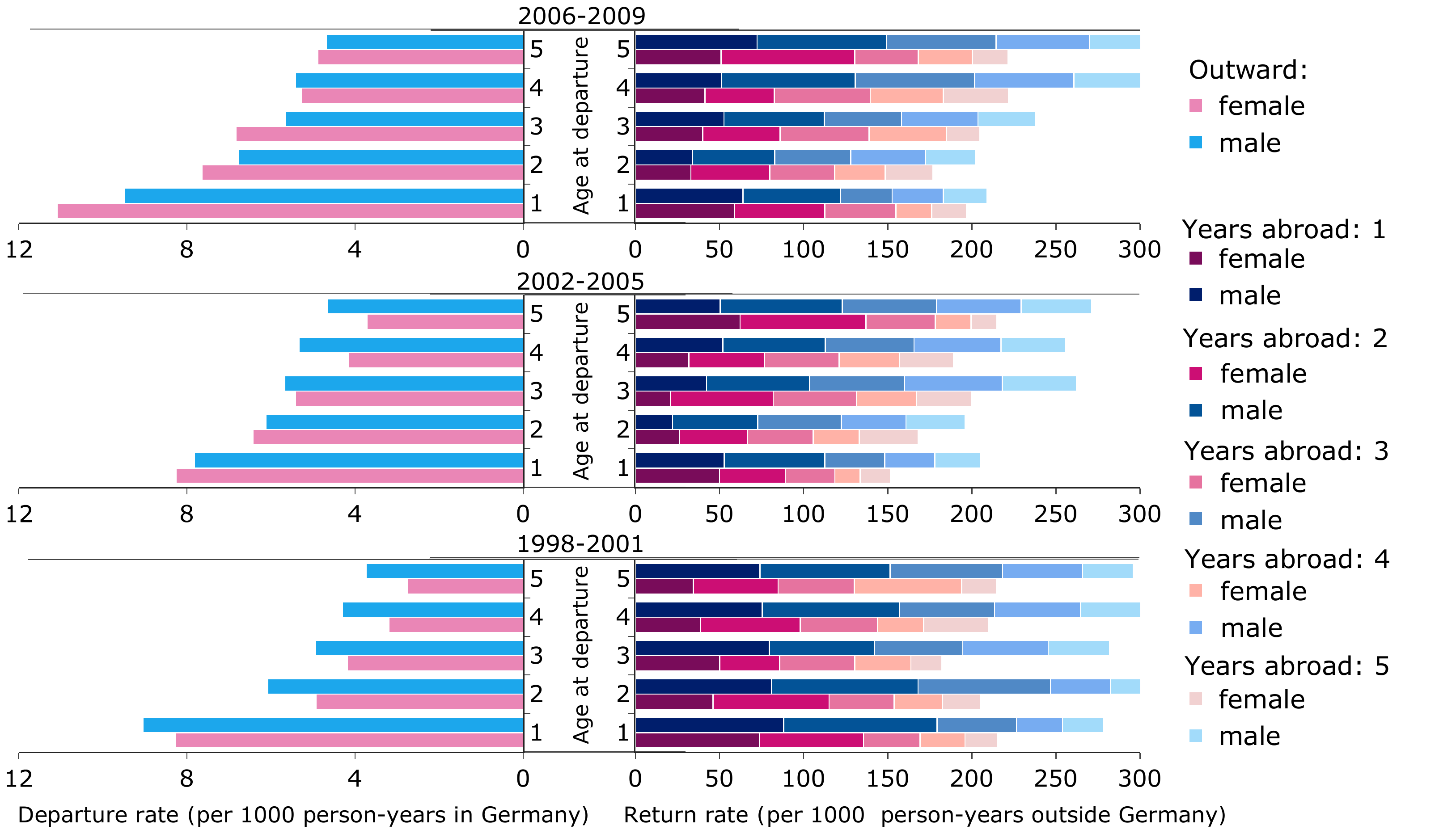}
		\caption{The rates of leaving Germany within first 5 years since first publication per 1000 person-years (left), and the rates of return to Germany within the first 5 years after departure per 1000 person-years (right).
		}
		\label{fig:4}
	\end{figure}

	
	Figure \ref{fig:4} illustrates the departure rates (left) and the return rates (right) per 1,000 person-years, disaggregated by cohort and gender. The academic age at departure is on the y-axis for both outward researchers and returnees. For returnees, the length of time away from Germany is also reported by the use of ombre colors. Taking the cohort 1998-2001 as an example, out of 1,000 researchers, around eight women and nine men with a German academic origin in this cohort moved abroad at academic age one. For every 1,000 outward researchers who left Germany at academic age one, around 215 women and 278 men had returned to Germany within five years. Among them, 74 women and 88 men had returned to Germany after one year, making the first year the most likely year of return for that cohort. In general, there was a slight but stable decline in the departure rates with academic age for all three cohorts. However, the most striking pattern is observed for the 2006-2009 cohort: the departure rates of female researchers exceeded those of male researchers for most ages, especially at academic ages one, two, and three. Specifically, we find that 11 out of 1,000 female researchers in this cohort left Germany at academic age one, while only nine out of 1,000 male researchers left Germany at that age. This result indicates that in this cohort, more female than male researchers chose to migrate early in their careers. Meanwhile, the return rates of the female researchers of all three cohorts were much lower than those of their male counterparts. This difference may be partly related to the longer average length of academic life for male returnees, as Figure \ref{fig:1} shows. Taken together, these results indicate that female outward researchers had a greater tendency than their male counterparts to remain abroad for longer periods or possibly to settle down in other countries, which may have exacerbated the gender disparities in the German science system. Thus, the findings suggest that out-migration trends may increase gender disparities within the German academic system unless further action is taken.
	
	We also observe that the return rates were generally higher for researchers who moved out of Germany in their later years, and tended to increase with academic age. This trend is more noticeable among male researchers and in the two latest cohorts. The more pronounced increase in return migration at later academic ages for men than for women suggests that there are structural processes that operate at specific moments of the academic life course, and that these processes could further extend the gender differences in German academia.
	
	\subsection{Gender composition of outward and return streams by discipline and cohort}
	\label{ss:discipline}
	
\begin{figure}[!htbp]
		\centering
		\includegraphics[width=0.95\textwidth]{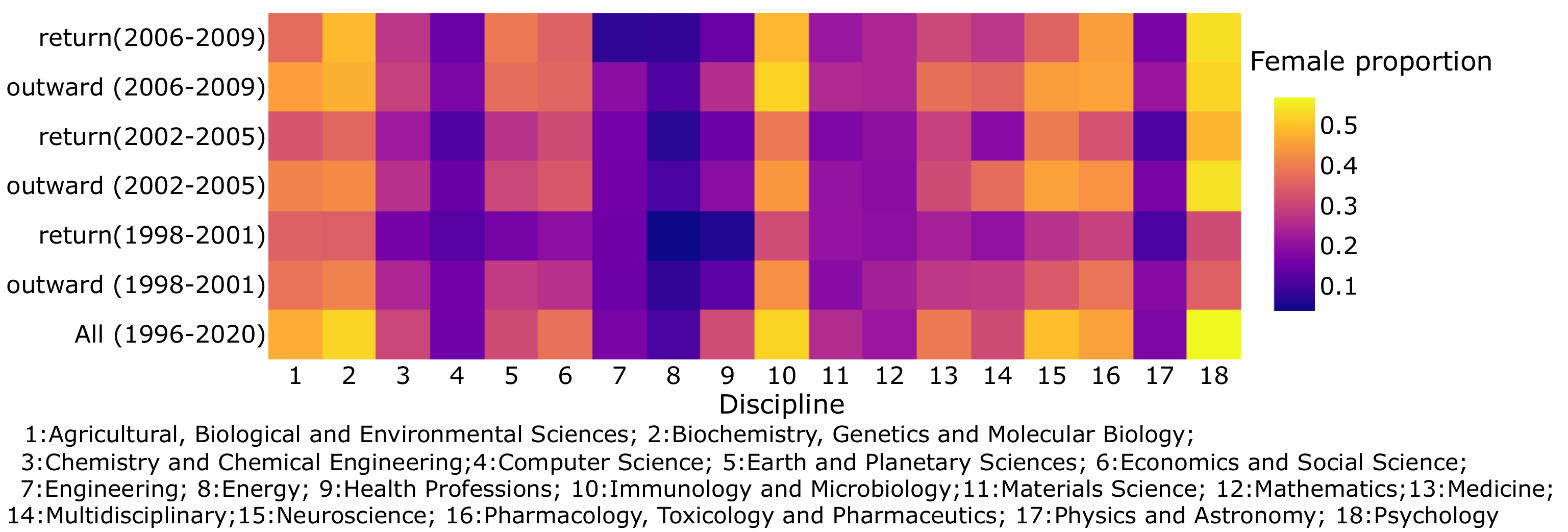}
		\caption{Proportion of female researchers in different groups by discipline and cohort}
		\label{fig:5}
	\end{figure}

	Considering that the male-to-female ratios of researchers vary across disciplines \citep{zhao2021international}, we take a further look at the gender disparities disaggregated by discipline for the three cohorts, as shown in Figure \ref{fig:5}. The colors in the heat map show that the representation of female researchers varies by discipline in the horizontal dimension, and by cohort in the vertical dimension. The bottom row of the map represents the overall proportion of female researchers in each discipline in Germany over the 1996-2020 period, as a baseline for comparing the variability in the representation of females among the researchers who left and returned over time.
	
	Compared to the baseline, almost all disciplines appear to be more male-dominated over time among both outward and returnee researchers, albeit to varying degrees. One exception is the field of mathematics (12), in which female researchers accounted for a higher proportion of each of these two migrant categories in the latest cohort (2006-2009), relative to the long-term pattern. Despite the lower representation of female researchers in both the outward and the returnee groups, for the majority of the disciplines, we see an increasing trend in the proportion of female researchers with each successive cohort, in line with our earlier discussion in Subsection \ref{ss:gender}.
	
	When comparing the categories of outward researchers and returnees in the same discipline and cohort, we observe that the proportion of female returnees was generally smaller than the proportion of female outward researchers. For example, when we look at the latest cohort of researchers in the field of energy, we find that the female proportion among returnees was much smaller than the female proportion among outward researchers. 
	The overall impression provided by these data is that most disciplines are experiencing rising gender disparities, in part because female scientists who leave Germany are less likely than their male counterparts to return. Despite significant efforts to increase gender equality in academia, gender disparities seem to remain substantial across disciplines.
	
	\subsection{Collaborative ties with Germany and rates of return} \label{ss:collaboration}
	
\begin{figure}
		\centering
		\includegraphics[width=0.7\textwidth]{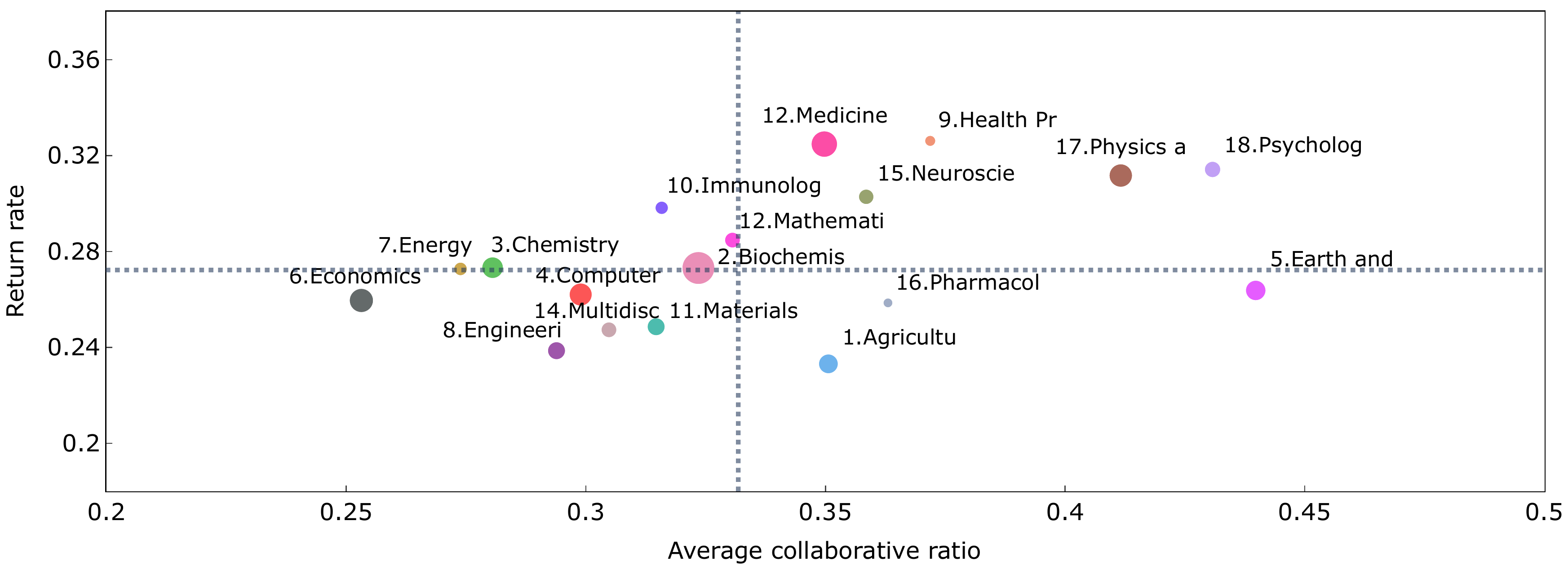}
		\caption{Return rates and collaborative ratios across disciplines}
		\label{fig:6}
	\end{figure}

	In this section, we examine the association between the levels of academic collaboration with Germany researchers maintained while abroad and the corresponding rates of return to Germany. Figure \ref{fig:6} shows a scatter plot of the return rates (y-axis) and the average collaborative ratios (x-axis) for each discipline. Note that the collaborative ratio is the fraction of publications of an outward researcher (during the period outside of Germany) with a German affiliation. The horizontal (vertical) line indicates the overall average return rates (average collaborative ratio) for outward researchers across all disciplines. The number of returnees for each discipline is represented by the size of the circles. Overall, the Pearson correlation coefficient between the collaborative ratio and the return rates is $0.45$, indicating a moderate positive association. Researchers in most health science and life science disciplines, including medicine, health professions, and psychology, were more likely to return to Germany than researchers in other disciplines, as indicated by the higher return rate over the average rate. When we look at the returnees' levels of academic collaboration with Germany while abroad, we see that health science returnees, as well as researchers in some physical science disciplines, like earth and planetary science, were more likely to maintain academic ties with Germany, as shown by collaborative ratios that exceed the mean values of $33\%$. Specifically, we observe that health science researchers maintained stronger collaborative ties with Germany, and were more likely to return; whereas researchers in STEM fields, who tended to leave Germany without maintaining as many collaborative ties, were less likely to return.
	
\begin{figure}
		\centering
		\includegraphics[width=0.98\textwidth]{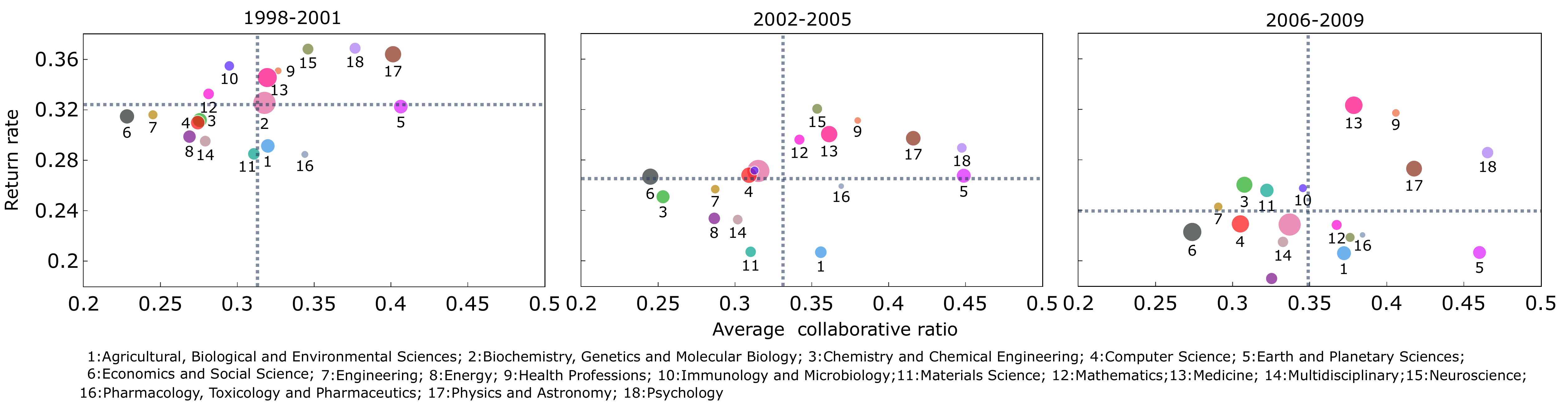}
		\caption{Return rates and collaborative ratios by discipline and cohort}
		\label{fig:7}
	\end{figure}

	Next, we look at the association between collaborative ties and return rates among outward researchers by cohort. The results disaggregated by cohort are shown in Figure \ref{fig:7}, with the average return rates and collaborative ratios in each cohort represented by the horizontal lines and the vertical lines, respectively. Our results show an overall decreasing trend in rates of return by cohort, but the left-truncation of the data complicates the reliable investigation of trends involving the first cohort. Despite the general trend, researchers in health professions and medicine were more likely to return than researchers in other disciplines. The collaborative ratios grew slowly but steadily with each cohort; thus, researchers in the latest cohort maintained relatively strong collaborative ties to Germany. Similar patterns can be observed separately for most disciplines.
	
	The correlations found between the collaborative ratios and return rates in the first two cohorts are in line with the overall pattern shown in Figure \ref{fig:6}, with Pearson correlation coefficients of 0.43 and 0.41, respectively. This association becomes much weaker (the correlation coefficient was 0.29) in the latest cohort, whose discipline averages appear to be scattered widely across the four quadrants. Between cohorts 2 and 3, neuroscience drops from quadrant 1 to quadrant 4, indicating a sharp decrease in return rates, despite an increase in academic links with Germany. Between cohorts 2 and 3, we see an increase in collaborative ratios among the outward researchers in the fields of chemistry and chemical engineering, accompanied by stable return rates. For most other disciplines, however, the return rates tended to decrease, as shown in Figure \ref{fig:7}.
	
	\section{Discussion and future directions}\label{sec4}
	
	As ``science brokers," researchers develop innovative ideas and make scientific contributions by combining information and resources in various domains using specialized skills and knowledge, which they acquire at different institutions and geographical locations \citep{Williams2007International}. International experience can play a substantial role in helping researchers accumulate knowledge, information, and capital, and can thus contribute to their scientific research and academic careers \citep{teichler_academic_2015,wang_temporal_2020}. Our previous study found that internationally mobile researchers accounted for over 16\% of the population of Scopus-published researchers who had affiliation ties to Germany over the 1996-2020 period \citep{zhao2021international}. We also observed that despite representing a minority in the German science system, mobile researchers make substantial contributions, as evidenced by the finding that compared to non-mover researchers in Germany, they have higher annual citation rates \citep{zhao2021international}. Because of their more nuanced trajectories and international experience, returnees can make important contributions to the German science system. Here, we have analyzed the return migration of researchers to Germany from several perspectives; i.e., by taking into account their disciplines, cohorts, genders, and levels of collaboration with Germany while abroad.
	
	Our quantitative results for Scopus-published researchers with ties to Germany provide further evidence to support previous findings. The results of our comprehensive analysis of emigration and return migration as two outcomes for researchers who left the German science system indicate that the age and gender compositions of outward researchers and returnee researchers differed from those of non-movers. The median age for returnee researchers was up to six years higher than that of non-movers, which suggests that there were substantial differences in their levels of experience. All three groups of researchers differentiated by their levels of experience, from early-career to senior, were heavily dominated by men. The ongoing gender disparities we found throughout the academic life cycle were in line with the findings of previous studies \citep{Orsolya2021Gender}. In particular, we observed that the publishing careers of male returnees were, on average, longer than those of other groups, with more than half of them being in their senior career stage.
	
	The countries receiving the largest flows of researchers from Germany were shown to have some of the highest return rates as well. However, we also found that of the large numbers of German researchers who moved to Switzerland, Sweden, Austria, and Australia, relatively small proportions returned to Germany. Three of these countries have linguistic, cultural, and geographic proximity to Germany. Moreover, they all have higher R\&D spending per GDP \citep{RDGDP} than the UK and the US (and three have higher R\&D spending than Germany), which has enabled them to succeed in attracting and retaining published researchers from Germany.
	
	Supporting the representation of female researchers in academia through equitable policies is imperative for Germany \citep{lutter2020there}, and for other countries \citep{morgan2021unequal}. The trajectories of internationally mobile female researchers is a particularly important dimension in evaluating a national science system. We analyzed the gender differences among outward researchers and returnees. Our results indicate that the gender disparities in the German science system tend to be intensified over cohorts. Consistent with evidence showing that the representation of female researchers in academia has rise over time \citep{huang_historical_2020}, we found that the proportion of female researchers has increased among both outward researchers and returnees across cohorts, taking into account the number of years between first publication, departure from Germany, and return to Germany. However, the proportion of female researchers among those who returned to Germany was lower than it was among those who left, which indicates that female outward researchers have a greater tendency than their male counterparts to live abroad for longer periods, or possibly to settle down in other countries. When we looked at the proportions of female researchers in the two subpopulations of interest disaggregated by cohort and discipline, we found that both the outward and the returnee subpopulations in most disciplines were more male-dominated than the overall population of researchers in that discipline, in line with the greater gender disparities observed among all German-affiliated migrant researchers in most disciplines \citep{zhao2021international}. These findings suggest that the gender imbalance in the German science system (with respect to scholars who started publishing in Germany) may be intensified by the subgroups who are returning to Germany being more male-dominated than the subgroups who are leaving Germany.
	
	Finally, we looked at the interplay between the degree to which researchers continued to collaborate with German institutions while abroad, and their corresponding return rates. The results showed a positive moderate association between collaboration and return rates across disciplines. After cohorts were introduced into the analysis, the return rates decreased with successive cohorts, while the collaborative ratios increased on average. In the fields of medicine, health professions, physics, and psychology, the likelihood of collaborating with Germany and of returning to Germany were both higher than the total averages. In contrast, researchers in the fields of engineering, computer science, and economics had both lower collaboration and lower return rates that the total average. To tackle the challenge of talent loss in STEM fields, and to attract and retain STEM researchers from abroad, Germany --which already has a large number of initiatives for international researchers, like GAIN and GSO-- would likely benefit from developing additional programs focused on STEM fields \citep{OECD_return_2015}.
	
	Our study has several limitations, which can be addressed only through ongoing work and additional efforts. Our bibliometric analysis was based on the higher quality signals for researchers who have higher publication rates. Therefore, the reliability of our findings may not be the same for all fields, given that their average publication rates vary (e.g., physics vs. history). Another limitation is that we could not analyze migration events that were not captured in the publication data. In addition, because of the possible differences between publication years and migration years, the temporal patterns of the data should be interpreted with caution.
	
	We recognize that bibliometric data, like other sources of big data, are not produced for use as research data, and are therefore susceptible to potential biases or errors. In our materials and methods, we outlined a series of pre-processing steps for systematically dealing with some of the data quality issues in our application context. Additional scientometrics research is needed to better identify the potential quality problems with bibliometric data, and to find systematic and effective remedies for addressing them.
	
	As well as contributing to the literature on the migration of researchers \citep{moed_bibliometric_2014,aman_does_2018, aman_transfer_2018, aref_demography_2019, Lovakov2019Biblio, robinson-garcia_many_2019, gonzalez2020scholarly,subbotin2020brain,EL2021Analyze} in the context of Germany \citep{netz_mobilised_2014, parey2017selection, zhao2021international}, more importantly, our research fills a critical gap in the research on the return migration of scholars, which is a novel subject in the bibliometric analysis of academic migration. This work, which represents a continuation of \cite{zhao2021international}, was aimed to provide a policy-relevant descriptive analysis of return migration among researchers by taking their levels of experience, gender, disciplines, and cohorts into account. Obtaining insights into researchers who have left Germany, including into their age, gender, and characteristics that could influence their potential return to Germany, is a key step towards understanding migration among scholars as a concept that is more nuanced than a one-off relocation event.
	
	A number of interesting questions still remain to be investigated, including the question of what personal and professional factors drive the international migration of researchers. Differences in levels of support for parenthood between Germany \citep{gangl2009motherhood, lutter2020there} and other countries \citep{morgan2021unequal} may have a bearing on some of the observed gender disparities. Combining different data sources could allow us to expand the analysis and examine other critical topics, like parenthood policies. Investigating the citation performance of outward and returnee researchers could provide us with additional insights into the individual-level consequences of scholars' migration decisions. In addition, the observed association between return migration and personal and professional factors, including disciplines and collaborative ties, can be further investigated with the aim of finding the mechanisms involved, such as the emergence of discipline-specific centers that are particularly attractive for migrant researchers. 
	
	\section*{Declarations}
	
	\bmhead{Availability of data and material}
	The bibliometric data used in this study is proprietary and cannot be released.
	Scopus data is owned and maintained by Elsevier. Our inferred migration events dataset is publicly available in a FigShare data repository \citep{zhao2022data}.
	
	\bmhead{Competing interests}
	The authors have no competing interests to declare that are relevant to the content of this article.
	
	\bmhead{Funding}
	This study has been funded by the German Academic Exchange Service (DAAD) with funds from the Federal Ministry of Education and Research (BMBF). This study has received access to the bibliometric data through the project ``Kompetenzzentrum Bibliometrie," and the authors acknowledge their funder BMBF (funding identification number 01PQ17001).
	
	\bmhead{Acknowledgments}
	This article is a substantially extended version of the paper \cite{zhao2021international} presented at the 18$^\text{th}$ International Conference on Scientometrics and Informetrics (ISSI 2021). The authors highly appreciate the comments from anonymous reviewers, the suggestions from Rezvaneh Rezapour, and the technical support from Tom Theile.

	
	\section*{Appendix: Mapping topics to disciplines}

	Table \ref{tab:mapping} provides the intermediate results from the topic model (30 topics) and how the topics are mapped into 17 disciplines (inspired by the All Science Journal Classification) based on their similarities.
	
	\clearpage
	
	\footnotesize
	\begin{longtable}{p{0.03\linewidth} p{0.57\linewidth}  p{0.3\linewidth}}
		\caption{Details of mapping 30 topics to 17 disciplines}
		\label{tab:mapping} \\
		\hline
		Topic & Most frequent keywords & Discipline result \\
		\hline
		01  & space, earth, wave, model, surface, field, datum, rock, seismic, structure & Earth and Planetary Sciences \\
		02  & material, property, surface, apply, structure, film, growth, metal, thin film, magnetic & Material Science \\
		03  & disease, cardiovascular, patient, clinical, function, heart, medicine, cardiac, lung & Medicine \\ 
		04  & infection, disease, clinical, immunology, virus, patient, vaccine, microbiology, transplantation, blood & Immunology and Microbiology  \\
		05  & ecology, population, evolution, forest, diversity, specie, genetic, biology, animal, conservation & Agricultural, Biological and Environmental Sciences \\ 
		06  & protein, molecular, gene, genetic, biology, cell, biological, expression, nature, mutation & Biochemistry, Genetics and Molecular Biology \\ 
		07  & model, simulation, engineering, flow, dynamic, design, numerical, structure, modeling, experimental & Engineering \\ 
		08  & network, note computer, subserie note artificial intelligence, note bioinformatic, ieee, information, model, datum, design, engineering & Computer Science \\ 
		09  & patient, surgery, clinical, treatment, cancer, outcome, disease, therapy, pediatric  & Medicine \\ 
		10  & plant, cell, metabolism, physiology, metabolic, stress, biology, response, enzyme, arabidopsis & Agricultural, Biological and Environmental Sciences \\ 
		11  & brain, neuroscience, rat, mouse, alzheimer disease, neurology, parkinson disease, disease, model, receptor & Neuroscience \\ 
		12  & laser, optical, engineering, spie society optical, optic, measurement, spectroscopy, fiber, pulse, apply & Physics and Astronomy  \\ 
		13  & management, health, economic, education, social, germany, development, policy,  review & Economics and Social Science \\ 
		14  & cell, cancer, expression, tumor, gene, molecular, stem, receptor, mouse, apoptosis & Biochemistry, Genetics and Molecular Biology \\
		15  & polymer, material, composite, engineering, property, surface, fiber, coating, application, technology & Material Sciences \\
		16  & nanoparticle, chemistry, surface, cell, membrane, chemical, physical, spectroscopy, microscopy, electrochemical & Chemistry and Chemical Engineering \\
		17  & chemistry, synthesis, chemical, structure, complex, reaction, organic, crystal, molecular,  & Chemistry and Chemical Engineering \\
		18  & cancer, patient, therapy, treatment, clinical, breast cancer, carcinoma, radiotherapy, tumor, oncology & Medicine \\
		19  & galaxy, star, astrophysical, cluster, society, xray, monthly notice royal astronomical, formation, general, evolution & Earth and Planetary Sciences \\
		20  & power, ieee, electronic, communication, technology, sensor, ieee transaction, measurement, device, application & Engineering \\
		21  & energy, process, technology, production, engineering, gas, atmospheric, environmental, water, development & Energy \\
		22  & drug, clinical, skin, exposure, risk, allergy, medicine, test, assessment,  & Pharmacology, Toxicology and Pharmaceutics \\
		23  & physics, physical review, letter, physical, energy, measurement, letter section, high,  nuclear, elementary particle, high energy & Physics and Astronomy \\
		24  & water, marine, environmental, new, sediment, climate, carbon, change, soil, ocean & Earth and Planetary Sciences \\
		25  & patient, child, disorder, cognitive, treatment, therapy, rehabilitation, physical, depression, pain & Psychology \\
		26  & model, theory, datum, mathematical, function, application, problem, dynamic, stochastic, time & Mathematics \\
		27  & surgery, bone, injury, treatment, fracture, tissue, trauma, surgical, clinical, implant & Medicine \\
		28  & soil, food, plant, apply, microbial, nutrition, activity, production, quality, microbiology & Agricultural, Biological and Environmental Sciences \\
		29  & imaging, image, magnetic resonance, mri, medicine, stroke, brain, radiology, ultrasound, compute tomography & Health Professions \\
		30  & plasma, physics, solar, nuclear, fusion, beam, ion, electron, radiation, nuclear instrument & Physics and Astronomy \\
		\hline
	\end{longtable}


\begin{thebibliography}{63}
	\providecommand{\natexlab}[1]{#1}
	\providecommand{\url}[1]{{#1}}
	\providecommand{\urlprefix}{URL }
	\providecommand{\doi}[1]{\url{https://doi.org/#1}}
	\providecommand{\eprint}[2][]{\url{#2}}
	\bibcommenthead
	
	\bibitem[{Abramitzky et~al(2019)Abramitzky, Boustan, and
		Eriksson}]{abramitzky2019new}
	Abramitzky R, Boustan L, Eriksson K (2019) To the new world and back again:
	Return migrants in the age of mass migration. ILR Review 72(2):300--322.
	\doi{10.1177/0019793917726981}
	
	\bibitem[{Ackers and Gill(2005)}]{ackers_attracting_2005}
	Ackers L, Gill B (2005) Attracting and retaining ‘early career’ researchers
	in english higher education institutions. Innovation: The European Journal of
	Social Science Research 18(3):277--299. \doi{10.1080/13511610500186649}
	
	\bibitem[{Aman(2018{\natexlab{a}})}]{aman_does_2018}
	Aman V (2018{\natexlab{a}}) Does the {Scopus} author {ID} suffice to track
	scientific international mobility? {A} case study based on {Leibniz}
	laureates. Scientometrics 117(2):705--720. \doi{10.1007/s11192-018-2895-3}
	
	\bibitem[{Aman(2018{\natexlab{b}})}]{aman_transfer_2018}
	Aman V (2018{\natexlab{b}}) Transfer of formal knowledge through international
	scientific mobility-introduction of a network-based bibliometric method. In:
	STI 2018 Conference Proceedings. Centre for Science and Technology Studies
	(CWTS), Leiden, pp 545--552, \doi{10.1162/qss_a_00028}
	
	\bibitem[{Andrey and Elena(2019)}]{Lovakov2019Biblio}
	Andrey L, Elena A (2019) Bibliometric analysis of publications from postsoviet
	countries in psychological journals in 1992–2017. Scientometrics
	119:1157–--1171. \doi{10.1007/s11192-019-03087-y}
	
	\bibitem[{Appelt et~al(2015)Appelt, van Beuzekom, Galindo-Rueda, and
		de~Pinho}]{appelt_chapter_2015}
	Appelt S, van Beuzekom B, Galindo-Rueda F, et~al (2015) Chapter 7 - which
	factors influence the international mobility of research scientists? In:
	Geuna A (ed) Global Mobility of Research Scientists. Academic Press, London,
	\doi{10.1016/B978-0-12-801396-0.00007-7}
	
	\bibitem[{Aref et~al(2019)Aref, Zagheni, and West}]{aref_demography_2019}
	Aref S, Zagheni E, West J (2019) The demography of the peripatetic researcher:
	{Evidence} on highly mobile scholars from the {Web} of {Science}. In:
	International {Conference} on {Social} {Informatics}. Springer, Cham, pp
	50--65, \doi{10.1007/978-3-030-34971-4_4}
	
	\bibitem[{Bardin(2016)}]{al_problem_2016}
	Bardin A (2016) The {Problem} of {Migration} in {German} {Academic}
	{Discourse}. Political Studies (6):183--188. \doi{10.17976/jpps/2016.06.13}
	
	\bibitem[{Baruffaldi and Landoni(2012)}]{baruffaldi_return_2012}
	Baruffaldi SH, Landoni P (2012) Return mobility and scientific productivity of
	researchers working abroad: The role of home country linkages. Research
	Policy 41(9):1655--1665. \doi{10.1016/j.respol.2012.04.005}
	
	\bibitem[{Bauder(2015)}]{bauder_international_2015}
	Bauder H (2015) The international mobility of academics: A labour market
	perspective. International Migration 53(1):83--96.
	\doi{https://doi.org/10.1111/j.1468-2435.2012.00783.x}
	
	\bibitem[{Blei(2012)}]{blei_probabilistic_2012}
	Blei DM (2012) Probabilistic topic models. Communications of the {ACM}
	55(4):77--84. \doi{10.1145/2133806.2133826}
	
	\bibitem[{Blei et~al(2003)Blei, Ng, and Jordan}]{blei_latent_2003}
	Blei DM, Ng AY, Jordan MI (2003) Latent dirichlet allocation. Journal of
	Machine Learning Research 3:993--1022. \doi{10.1162/jmlr.2003.3.4-5.993}
	
	\bibitem[{Burnham(2006)}]{burnham_scopus_2006}
	Burnham JF (2006) Scopus database: a review. Biomedical Digital Libraries
	3(1):1. \doi{10.1186/1742-5581-3-1}
	
	\bibitem[{Conchi and Michels(2014)}]{conchi_scientific_2014}
	Conchi S, Michels C (2014) Scientific mobility: An analysis of {Germany},
	{Austria}, {France} and {Great Britain}. Working paper, Fraunhofer {ISI}
	Discussion Papers Innovation Systems and Policy Analysis,
	\urlprefix\url{https://www.econstor.eu/handle/10419/94371}
	
	\bibitem[{Dahal et~al(2019)Dahal, Kumar, and Li}]{dahal_topic_2019}
	Dahal B, Kumar SAP, Li Z (2019) Topic modeling and sentiment analysis of global
	climate change tweets. Social Network Analysis and Mining 9(1):24.
	\doi{10.1007/s13278-019-0568-8}
	
	\bibitem[{D'Angelo and van Eck(2020)}]{dangelo_collecting_2020}
	D'Angelo CA, van Eck NJ (2020) Collecting large-scale publication data at the
	level of individual researchers: A practical proposal for author name
	disambiguation. Scientometrics 123:883--907. \doi{10.1007/s11192-020-03410-y}
	
	\bibitem[{Dorling et~al(2006-09)Dorling, Barford, and
		Newman}]{dorling_worldmapper_2006}
	Dorling D, Barford A, Newman M (2006-09) Worldmapper: The world as you've never
	seen it before. {IEEE} Transactions on Visualization and Computer Graphics
	12(5):757--764. \doi{10.1109/TVCG.2006.202}
	
	\bibitem[{Düvell(2019)}]{duvell_germany_nodate}
	Düvell F (2019) Germany: {Selected} {Migration} {Policies}. Tech. Rep. SEO
	Report No. 2019-24, Dutch Ministry of Foreign Affairs,
	\href{https://25cjk227xfsu3mkyfg1m9xb7-wpengine.netdna-ssl.com/wp-content/uploads/2019/01/Annex_E_Germany.pdf}{https://25cjk227xfsu3mkyfg1m9xb7-wpengine.netdna-ssl.com/wp-content/uploads/2019/01/Annex_E_Germany.pdf}
	
	\bibitem[{El-Ouahi et~al(2021)El-Ouahi, Nicolas, and Rodrigo}]{EL2021Analyze}
	El-Ouahi J, Nicolas RG, Rodrigo C (2021) Analyzing scientific mobility and
	collaboration in the {Middle East} and {North Africa}. Quantitative Science
	Studies pp 1--30. \doi{10.1007/s11192-019-03087-y}
	
	\bibitem[{Elsevier(2020)}]{scopus_coverage}
	Elsevier (2020) Scopus content coverage guide.
	\urlprefix\url{https://www.elsevier.com/solutions/scopus/how-scopus-works/content}
	
	\bibitem[{Eule(2016)}]{eule_inside_2016}
	Eule TG (2016) Inside {Immigration} {Law}: {Migration} {Management} and
	{Policy} {Application} in {Germany}. Routledge, Oxfordshire,
	\doi{10.4324/9781315588728}
	
	\bibitem[{Fernández-Zubieta et~al(2015)Fernández-Zubieta, Geuna, and
		Lawson}]{fernandez-zubieta_chapter_2015}
	Fernández-Zubieta A, Geuna A, Lawson C (2015) Chapter 1 - what do we know of
	the mobility of research scientists and impact on scientific production. In:
	Geuna A (ed) Global Mobility of Research Scientists. Academic Press, London,
	p 1--33, \doi{10.1016/B978-0-12-801396-0.00001-6}
	
	\bibitem[{Franzoni et~al(2014)Franzoni, Scellato, and
		Stephan}]{franzoni2014mover}
	Franzoni C, Scellato G, Stephan P (2014) The mover’s advantage: The superior
	performance of migrant scientists. Economics Letters 122(1):89--93.
	\doi{10.1016/j.econlet.2013.10.040}
	
	\bibitem[{Franzoni et~al(2015)Franzoni, Scellato, and
		Stephan}]{franzoni_chapter_2015}
	Franzoni C, Scellato G, Stephan P (2015) Chapter 2 - international mobility of
	research scientists: Lessons from {GlobSci}. In: Geuna A (ed) Global Mobility
	of Research Scientists. Academic Press, London, p 35--65,
	\doi{10.1016/B978-0-12-801396-0.00002-8}
	
	\bibitem[{Gangl and Ziefle(2009)}]{gangl2009motherhood}
	Gangl M, Ziefle A (2009) Motherhood, labor force behavior, and women’s
	careers: An empirical assessment of the wage penalty for motherhood in
	{Britain}, {Germany}, and the {United States}. Demography 46(2):341--369.
	\doi{10.1353/dem.0.0056}
	
	\bibitem[{Gerlach et~al(2018)Gerlach, Peixoto, and
		Altmann}]{gerlach_network_2018}
	Gerlach M, Peixoto TP, Altmann EG (2018) A network approach to topic models.
	Science Advances 4(7):eaaq1360. \doi{10.1126/sciadv.aaq1360}
	
	\bibitem[{Guthrie et~al(2017)Guthrie, Lichten, Harte, Parks, and
		Wooding}]{guthrie_international_2017}
	Guthrie S, Lichten CA, Harte E, et~al (2017) International mobility of
	researchers: A survey of researchers in the {UK}. RAND Corporation, Santa
	Monica,
	\href{https://royalsociety.org/~/media/policy/projects/international-mobility/researcher-mobility-report-survey-academics-uk.pdf}{https://royalsociety.org/~/media/policy/projects/international-mobility/researcher-mobility-report-survey-academics-uk.pdf}
	
	\bibitem[{Houle et~al(2009)Houle, Holt, Gillespie, Freedman, and
		Reyes}]{houle_use_2009}
	Houle B, Holt J, Gillespie C, et~al (2009) Use of density-equalizing cartograms
	to visualize trends and disparities in state-specific prevalence of obesity:
	1996–2006. American Journal of Public Health 99(2):308--312.
	\doi{10.2105/AJPH.2008.138750}
	
	\bibitem[{Huang et~al(2020)Huang, Gates, Sinatra, and
		Barabási}]{huang_historical_2020}
	Huang J, Gates AJ, Sinatra R, et~al (2020) Historical comparison of gender
	inequality in scientific careers across countries and disciplines.
	Proceedings of the National Academy of Sciences 117(9):4609--4616.
	\doi{10.1073/pnas.1914221117}
	
	\bibitem[{Kawashima and Tomizawa(2015)}]{kawashima_accuracy_2015}
	Kawashima H, Tomizawa H (2015) Accuracy evaluation of {Scopus} {Author} {ID}
	based on the largest funding database in {Japan}. Scientometrics
	103(3):1061--1071. \doi{10.1007/s11192-015-1580-z}
	
	\bibitem[{Larivi{\`e}re et~al(2013)Larivi{\`e}re, Ni, Gingras, Cronin, and
		Sugimoto}]{lariviere2013bibliometrics}
	Larivi{\`e}re V, Ni C, Gingras Y, et~al (2013) Bibliometrics: Global gender
	disparities in science. Nature 504(7479):211. \doi{10.1007/s11192-015-1580-z}
	
	\bibitem[{Lutter and Schr{\"o}der(2020)}]{lutter2020there}
	Lutter M, Schr{\"o}der M (2020) Is there a motherhood penalty in academia? the
	gendered effect of children on academic publications in {German} sociology.
	European Sociological Review 36(3):442--459. \doi{10.1093/esr/jcz067}
	
	\bibitem[{Macaluso et~al(2016)Macaluso, Larivière, Sugimoto, and
		Sugimoto}]{macaluso_is_2016}
	Macaluso B, Larivière V, Sugimoto T, et~al (2016) Is science built on the
	shoulders of women? a study of gender differences in contributorship.
	Academic Medicine 91(8):1136--1142. \doi{10.1097/ACM.0000000000001261}
	
	\bibitem[{Melin and Janson(2006)}]{melin_what_2006}
	Melin G, Janson K (2006) What skills and knowledge should a {PhD} have?
	changing preconditions for {PhD} education and post doc work. In: The
	Formative Years of Scholars, Wenner-Gren International Series, vol~83.
	Portland Press, London, p 105--118,
	\href{https://unike.au.dk/fileadmin/www.unike.au.dk/What_Skills___Knowledge_Should_a_PhD_Have.pdf}{https://unike.au.dk/fileadmin/www.unike.au.dk/What_Skills___Knowledge_Should_a_PhD_Have.pdf}
	
	\bibitem[{Miranda-Gonz{\'a}lez et~al(2020)Miranda-Gonz{\'a}lez, Aref, Theile,
		and Zagheni}]{gonzalez2020scholarly}
	Miranda-Gonz{\'a}lez A, Aref S, Theile T, et~al (2020) Scholarly migration
	within {Mexico}: {Analyzing} internal migration among researchers using
	{Scopus} longitudinal bibliometric data. EPJ Data Science 9:1--26.
	\doi{10.1140/epjds/s13688-020-00252-9}
	
	\bibitem[{Moed and Halevi(2014)}]{moed_bibliometric_2014}
	Moed HF, Halevi G (2014) A bibliometric approach to tracking international
	scientific migration. Scientometrics 101(3):1987--2001.
	\doi{10.1007/s11192-014-1307-6}
	
	\bibitem[{Moed et~al(2013)Moed, Aisati, and Plume}]{moed_studying_2013}
	Moed HF, Aisati M, Plume A (2013) Studying scientific migration in {Scopus}.
	Scientometrics 94(3):929--942. \doi{10.1007/s11192-012-0783-9}
	
	\bibitem[{Mongeon and Paul-Hus(2016)}]{mongeon_journal_2016}
	Mongeon P, Paul-Hus A (2016) The journal coverage of {Web} of {Science} and
	{Scopus}: A comparative analysis. Scientometrics 106(1):213--228.
	\doi{10.1007/s11192-015-1765-5}
	
	\bibitem[{Morgan et~al(2021)Morgan, Way, Hoefer, Larremore, Galesic, and
		Clauset}]{morgan2021unequal}
	Morgan AC, Way SF, Hoefer MJ, et~al (2021) The unequal impact of parenthood in
	academia. Science Advances 7(9):eabd1996. \doi{10.1126/sciadv.abd1996}
	
	\bibitem[{Netz and Jaksztat(2014)}]{netz_mobilised_2014}
	Netz N, Jaksztat S (2014) Mobilised by mobility? determinants of international
	mobility plans among doctoral candidates in {Germany}.
	\doi{10.1108/S1479-362820140000011009}, archive Location: world {ISBN}:
	9781783508532 {ISSN}: 1479-3628 Publisher: Emerald Group Publishing Limited
	
	\bibitem[{Netz and Jaksztat(2017)}]{netz2017explaining}
	Netz N, Jaksztat S (2017) Explaining scientists’ plans for international
	mobility from a life course perspective. Research in Higher Education
	58(5):497--519. \doi{10.1007/s11162-016-9438-7}
	
	\bibitem[{OECD(2008)}]{oecd_global_2008}
	OECD (2008) The global competition for talent: mobility of the highly skilled.
	{OECD Publishing}, Paris,
	\href{https://www.oecd-ilibrary.org/employment/the-global-competition-for-talent_9789264047754-en}{https://www.oecd-ilibrary.org/employment/the-global-competition-for-talent_9789264047754-en}
	
	\bibitem[{OECD(2015)}]{OECD_return_2015}
	OECD (2015) Talent Abroad: A Review of German Emigrants. OECD Publishing,
	Paris, \doi{10.1787/9789264231702-en}
	
	\bibitem[{OECD(2021)}]{RDGDP}
	OECD (2021) Gross domestic spending on {R}\&{D} (indicator).
	\urlprefix\url{https://data.oecd.org/rd/gross-domestic-spending-on-r-d.htm}
	
	\bibitem[{Parey et~al(2017)Parey, Ruhose, Waldinger, and
		Netz}]{parey2017selection}
	Parey M, Ruhose J, Waldinger F, et~al (2017) The selection of high-skilled
	emigrants. Review of Economics and Statistics 99(5):776--792.
	\doi{10.1162/REST_a_00687}
	
	\bibitem[{Paturi and Loktev(2020)}]{loktev2020}
	Paturi M, Loktev A (2020) The best gets better: {Scopus} data quality,
	measured, \urlprefix\url{https://www.brighttalk.com/webcast/13819/456949}
	
	\bibitem[{Pedregosa et~al(2011)Pedregosa, Varoquaux, Gramfort, Michel, Thirion,
		Grisel, Blondel, Prettenhofer, Weiss, and
		Dubourg}]{pedregosa_scikit-learn_2011}
	Pedregosa F, Varoquaux G, Gramfort A, et~al (2011) Scikit-learn: {Machine}
	learning in {Python}. the Journal of machine Learning research 12:2825--2830.
	\href{https://www.jmlr.org/papers/volume12/pedregosa11a/pedregosa11a.pdf}{https://www.jmlr.org/papers/volume12/pedregosa11a/pedregosa11a.pdf}
	
	\bibitem[{Pritchard et~al(2000)Pritchard, Stephens, and
		Donnelly}]{pritchard_inference_2000}
	Pritchard JK, Stephens M, Donnelly P (2000) Inference of population structure
	using multilocus genotype data. Genetics 155(2):945--959.
	\urlprefix\url{https://www.genetics.org/content/155/2/945}
	
	\bibitem[{Reilly et~al(2005)Reilly, Armstrong, Dorosty, Emmett, Ness, Rogers,
		Steer, and Sherriff}]{reilly_early_2005}
	Reilly JJ, Armstrong J, Dorosty AR, et~al (2005) Early life risk factors for
	obesity in childhood: cohort study. {BMJ} 330(7504):1357.
	\doi{10.1136/bmj.38470.670903.E0}
	
	\bibitem[{Robinson-Garcia et~al(2019)Robinson-Garcia, Sugimoto, Murray,
		Yegros-Yegros, Larivière, and Costas}]{robinson-garcia_many_2019}
	Robinson-Garcia N, Sugimoto CR, Murray D, et~al (2019) The many faces of
	mobility: Using bibliometric data to measure the movement of scientists.
	Journal of Informetrics 13(1):50--63. \doi{10.1016/j.joi.2018.11.002},
	\urlprefix\url{http://www.sciencedirect.com/science/article/pii/S1751157718300865}
	
	\bibitem[{Rothman(2012)}]{rothman_epidemiology_2012}
	Rothman KJ (2012) Epidemiology: An Introduction. Oxford University Press,
	Oxford, \doi{10.1093/aje/kwf028}
	
	\bibitem[{Schiller and Cordes(2016)}]{schiller_measuring_nodate}
	Schiller D, Cordes A (2016) Measuring researcher mobility. In: OECD Blue Sky
	Forum, p~24,
	\href{https://www.oecd.org/sti/062\%20-\%20Schiller-Cordes-Researcher-Mobility-final.pdf}{https://www.oecd.org/sti/062\%20-\%20Schiller-Cordes-Researcher-Mobility-final.pdf}
	
	\bibitem[{Subbotin and Aref(2021)}]{subbotin2020brain}
	Subbotin A, Aref S (2021) Brain drain and brain gain in {Russia}: {Analyzing}
	international mobility of researchers by discipline using {Scopus}
	bibliometric data 1996-2020. Scientometrics 126(9):7875--7900.
	\doi{10.1007/s11192-021-04091-x}
	
	\bibitem[{Teichler(2015)}]{teichler_academic_2015}
	Teichler U (2015) Academic mobility and migration: What we know and what we do
	not know. European Review 23:S6--S37. \doi{10.1017/S1062798714000787}
	
	\bibitem[{Tokunaga and Makoto(1994)}]{tfidf}
	Tokunaga T, Makoto I (1994) Text categorization based on weighted inverse
	document frequency. In: Special Interest Groups and Information Process
	Society of Japan (SIG-IPSJ), pp 33--39, \doi{10.1.1.49.7015}
	
	\bibitem[{Turney(2019)}]{ResearchinSwitzerland}
	Turney A (2019) Opportunities for foreign researchers in {Switzerland}.
	\urlprefix\url{https://www.academics.com/guide/research-switzerland}
	
	\bibitem[{V{\'a}s{\'a}rhelyi et~al(2021)V{\'a}s{\'a}rhelyi, Zakhlebin,
		Milojevi{\'c}, and Horv{\'a}t}]{Orsolya2021Gender}
	V{\'a}s{\'a}rhelyi O, Zakhlebin I, Milojevi{\'c} S, et~al (2021) Gender
	inequities in the online dissemination of scholars’ work. Proceedings of
	the National Academy of Sciences 118(39). \doi{10.1073/pnas.2102945118}
	
	\bibitem[{Wang(2020)}]{wang_temporal_2020}
	Wang B (2020) A temporal gaze towards academic migration: Everyday times,
	lifetimes and temporal strategies amongst early career {Chinese} academic
	returnees. Time \& Society 29(1):166--186. \doi{10.1177/0961463X19873806},
	publisher: {SAGE} Publications Ltd
	
	\bibitem[{Weert(2013)}]{weert_support_2013}
	Weert Ed (2013) Support for continued data collection and analysis concerning
	mobility patterns and career paths of researchers. European Commission,
	Research Directorate-General. Directorate B - European Research Area,
	Brussels,
	\href{https://ris.utwente.nl/ws/portalfiles/portal/5142100/Weert_MORE_project_Support_for_conitued_data.pdf}{https://ris.utwente.nl/ws/portalfiles/portal/5142100/Weert_MORE_project_Support_for_conitued_data.pdf}
	
	\bibitem[{Williams(2007)}]{Williams2007International}
	Williams AM (2007) International labour migration and tacit knowledge
	transactions: a multi‐level perspective. Global Networks 7(1):29--50.
	\doi{10.1111/j.1471-0374.2006.00155.x}
	
	\bibitem[{Zhao et~al(2021)Zhao, Aref, Zagheni, and
		Stecklov}]{zhao2021international}
	Zhao X, Aref S, Zagheni E, et~al (2021) International migration in academia and
	citation performance: An analysis of {German}-affiliated researchers by
	gender and discipline using {Scopus} publications 1996-2020. In: Gl{\"a}nzel
	W, Heeffer S, Chi PS, et~al (eds) Proceedings of the 18th International
	Conference on Scientometrics and Informetrics. ISSI, Leuven, p 1369--1380,
	\href{https://kuleuven.app.box.com/s/kdhn54ndlmwtil3s4aaxmotl9fv9s329}{https://kuleuven.app.box.com/s/kdhn54ndlmwtil3s4aaxmotl9fv9s329}
	
	\bibitem[{Zhao et~al(2022)Zhao, Aref, Zagheni, and Stecklov}]{zhao2022data}
	Zhao X, Aref S, Zagheni E, et~al (2022) Dataset of international migration
	among {German}-affiliated researchers in {Scopus} over 1996-2020.
	\emph{FigShare}, \doi{10.6084/m9.figshare.18433139}
	
	\bibitem[{Zippel(2017)}]{zippel_women_2017}
	Zippel K (2017) Women in Global Science: Advancing Academic Careers through
	International Collaboration. Stanford University Press, California,
	\doi{10.1515/9781503601505}
	
	\end{thebibliography}
\end{document}